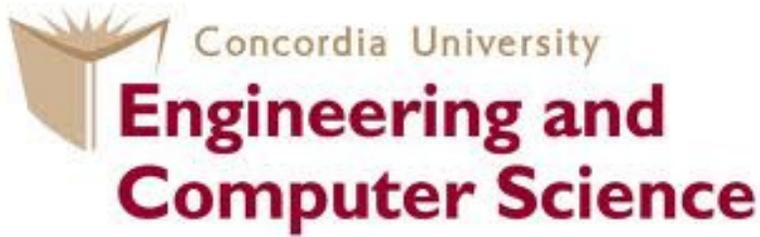
Concordia University
# Engineering and Computer Science

# Toward Recovering Complete SRS for Softbody Simulation System and a Sample Application

## A Team 4 SOEN6481-W13 Project Report

| | |
|---|---|
| **Oualid El Halimi** | **o_elhali@encs.concordia.ca** |
| **Peyman Derafshkavian** | **p_derafs@encs.concordia.ca** |
| **Abdulrhman Albeladi** | **a_albela@encs.concordia.ca** |
| **Faisal Alrashdi** | **f_alrash@encs.concordia.ca** |

**Supervised by:**

Dr. Serguei A. Mokhov

# Table of Contents

## List of Terms, Acronyms and Abbreviations

| Term / Acronym / Abbreviation | Expansion / Description / Definition |
|---|---|
| System | The Soft-body system |
| User | An end user who interacts with the Soft-body system |
| Elastic Object | An object that can be deformed. It could be a particle, spring, face, or integrator |
| Forces | Forces applied on the object under simulation. Forces could be: External, Gravity, Mouse, Spring, Pressure, or Collisions detection forces. |
| Deformation | Change in the shape of the object due to the forces applied. In Soft-body system, deformation is based on the realistic physical consistency of tissues and the laws of the established physics. |
| View Space | The dynamic layout that the Soft-body system provide to end users to set its content by dragging objects to it |
| Windows | The operating system of Microsoft Company |
| OS X | The operating system by Apple Inc |
| Linux | An open source operating system |
| Acrobat Reader | The software from Adobe company for PDF file processing |
| Simulation | Refers to the act of using a computer to see our tended results |
| Jelly Fish | An extension of an elastic object |
| OpenGL | Open Graphics Library |
| XML | Extensible Markup language |
| SRS | System Requirements Specification |
| UI | User Interface |
| MVC | Model View Controller design pattern |
| API | Application Programming Interface |
| LAN | Local Area Network |
| AHP | Analytic Hierarchy Process |
| SRS | System Requirement Specification |
| SSD | System Sequence Diagram |

Table 1 List of Terms, Acronyms and Abbreviations

# List of Tables

# List of Figures

# 1. Introduction

This document gathers high-level users' requirements and describes the system features. It provides a detailed explanation of the main functionalities of the system with a more emphasis on the stakeholders needs and wants. Indeed, the document goes through design constraints that may restrict various aspects of the design and implementation.

## 1.1 Purpose of the Vision Document

The following document provides an in-depth description of the Dynamic Deformation of Uniform Elastic Two-Layer Objects project [1], including an explanation of the stakeholders and the problems that the project will help solve for its clients.

## 1.2 Product Overview

The system to be designed simulates an object with extension of a Jelly Fish. It shows how it reacts to different actions and forces applied on it. This helps users to observe object deformation. The system analyzes all actions and applies the necessary reactions on the object.

# 2. User Description

This project is a research with regard to simulation of deformation of uniform elastic two-layer objects. As a consequence, this is not a commercial market driven project and the users of this project are limited to researchers including Instructors and Students.

## 2.1 User/Market Demographics

This project can target graphic and multimedia market; more precisely we can say that this project could be a foundation for developing graphical games.

## 2.2 User Profiles

The users of this project are mainly expert users including as below:

- Students: they need a friendly interface that simulates a Jelly Fish. They also need to apply different experiments on it to observe its reaction, and then print out a report to analyze the study results and submit it to their instructors. Furthermore, they can save any simulation so they can continue later on it.

- Instructors: they need a sophisticated system through which they can perform various studies to come up with new innovations. They also need auto analysis and results so it helps them to reach conclusions easily. Like students, they need to save the simulation.

Both users may not reach the desired conclusions if they do not have adequate training.

## 2.3 User Environment

Users can use the system individually or in groups of different sizes for completing their experiments. An experiment has no precise simulation time and finishes under user's wish.

The software can be deployed in any platform. It needs to be integrated with Adobe Reader to save or print reports. The simulations can be saved as CVS or XML files.

## 2.4 Key User Needs

Users need to perceive the deformation of objects in this system. The system shall provide the simulation with details in good graphic display mode in real time.

## 2.5 Alternatives and Competition

This system is a research in simulation of deformation of objects. As a consequence it does not tend to have any alternatives or competitors.

## 3. Product Overview

This project provides a simulation environment for the user to perceive and analyze two-layer objects deformation.

## 3.1 Product Perspective

This simulation system needs Windows OS or OSX or Linux as platforms and Acrobat Reader for providing report.

## 3.2 Product Position Statement

The below chart demonstrates the system position statement:

| | |
|---|---|
| For | Researchers including Instructors and Students |
| Who | Want to do experiments with simulation of two layer objects deformation |
| The System | Is a simulation software |
| That | Simulates of two layer objects deformation |
| Unlike | Not Applicable |
| Our Product | Saves any simulation for future use. |

## 3.3 Summary of Capabilities

The below chart describes capabilities of the system:

| Customer Benefit | Supporting Features |
|---|---|
| It is easy to learn how to use the software | Online help, tutorial and intuitive user interface |
| It is easy to use the software | Friendly user interface |
| Results are reliable | Real Time simulation |
| It is usable for all users with different backgrounds | Different software modes<br><br>Online help system provides sufficient guidance for novice users |

## 3.4 Assumptions and Dependencies

This system needs Windows OS or OSX or Linux as the platform; if this platform changes the vision document needs to be updated. It also needs Adobe Reader to save or print reports.

## 3.5 Cost and Pricing

This project is a research and does not aim any markets; therefore, no cost and pricing refers to it.

## 4. Feature Attributes

This system provides a simulation environment to see deformation of two-layer objects.

## 5. Product Features

Product features including Functional and Non-Functional Requirements are listed as below:

- Software objects shall be squashed and stretched according to external & internal forces applied on them.
- The two-layer elastic computer generated object provides more accurate modeling based on the main feature of human tissue.
- Deformation is based on the realistic physical consistency of tissues and the laws of the established physics.
- Program shall be easy in implementation, convenient to re-use, and give best elastic body behavior at the minimum cost.
- Users shall be able to interact with the software body in real-time
- The collision detection & response must be handled correctly.
- The elastic object simulation has to be accurate
- The three-dimensional elastic object consists of the same elements of 2D, such as particles, springs, and faces, but extended to z-axis.
- A uniform modeling technique has to be used for surface refinement to generate approximations to curves and surfaces of a sphere.
- The system shall be efficient, accurate, and stable.
- An elastic object consists of a particle, spring, face, and integrator.
- The system's graphical user interface provides the user an interactive environment.
- The system's graphical user interface allows the user to drag the object.
- The user shall set his view space and choose the object type:

- o One-dimensional.

- o Two-dimensional.

- o Three-dimensional

- The end user shall choose the integrator type:

  - o Eurler

  - o Midpoint

  - o Runge Kutta

- The end user shall set up the springs:

  - o Stiffness variables

  - o Damping variables

  - o Pressure

- The system shall create an elastic object and add particles, springs and faces to the layout.

- The system shall display the object with:

  - o The new position

  - o Velocity

  - o Deformed shape

- The system integrator shall be able to update the particles velocity and new position based on:

  - o Accumulated forces

  - o External forces

  - o Gravity forces

  - o Mouse forces

  - o Spring forces

  - o Pressure forces

  - o Collisions detection

- The system updates the object position upon the mouse position.

- The system shall allow Soft-body objects to be attached to other objects.

- The system shall promote interactivity through haptic devices with the Soft-body feedbacks.

- The user shall save a state dump at any time in a CVS or XML file.

- The user can reload functionality from a saved XML file for the following reasons:

  o Simulation

  o Replay

  o Debugging purposes.

- The XML file could be imported into:

  o Relational database.

  o Excel spreadsheet

- The system allows for multiple rendering back ends.

- The system allows stereoscopic effects.

# 6. Other Product Requirements

- Stability: the simulation system shall be stable.

- Re-usability: The design of this simulation system is based on well-known software design pattern Model-View-Controller that ease system's reusability.

- Portability: the system's source code shall be deployed under different platforms and build systems.

- The system uses graphical libraries like OpenGL, GLSL, GLUI, Direct X

- Extendibility: The system allows for extension by sub-classed applications.

- Readability: The code should be well-structured, commented, documented, use persistent naming and coding conventions and the API.

# 7. Documentation Requirements

## 7.1 Applicable Standards

The system should follow the international Organization for Standardization ISO9000 standards to ensure quality management needs and meet the wants of customers and stakeholders. Also it should follow ISO29127: 1988 standard to standardize User Documentation and cover information for consumer software packages. It needs Windows OS or OSX or Linux platform.

## 7.2 System Requirements
The system requires a minimum of 4 GB memory, and requires no Internet connection to run.

## 7.3 Licensing, Security, and Installation
This system is licensed for Concordia University of Montreal Canada and all rights reserved for its producer Miao Song.

## 7.4 Performance Requirements
The system shall provide real time response to users while they are interacting with it. It shall simulate all the changes occurring to the object.

# 8. Documentation Requirements

## 8.1 User Manual
- The user manual shall detail the minimum system requirements.
- It shall describe the use of the system.
- It shall list and describe the system's features.
- It shall be available online.

## 8.2 Online Help
- Online help shall be available 24/7.
- It shall be available for each function with demo.

## 8.3 Installation Guides, Configuration, and Read Me Files
- Installation guide shall show how to install the system with pictures step by step.
- Read me file shall list new features.
- It also shall list common troubleshooting and workaround.

## 8.4 Labeling and Packaging
Concordia University logo shall be shown in all documents.

# 9. Supplementary Specification

## 9.1 Introduction

Some user requirements like security and reliability are not being captured in the use case model. Therefore, Supplementary Document lists these requirements in an easy and organized manner to fully define the system-to-be functionalities. It includes the requirement definitions agreed upon, quality goals, and design constraints. We can find multiple quality goals like performance, reliability, supportability as well as usability. This document can also serve to identify system constraints such as used platform, used OpenGL libraries, performance issues, and interoperability with existing system libraries. Indeed, the Supplementary Specification complements the use case specification to produce a full set of the Soft-body system requirements.

### 9.1.1 Purpose

The purpose of the Supplementary Document is to define requirements of the Soft-body system that are not captured in the use cases. This document lists these requirements as a reference for team members to use throughout the software development process.

### 9.1.2 Scope

This Supplementary Specification applies to the Soft-body system developed for the department of Science & Software Engineering at Concordia University. The application is supervised by Dr. Serguei A. Mokhov for the system Requirements Specification course.

The Soft-body system should simulate real objects deformation based on the realistic physical consistency of tissues and the laws of the established physics. It allows the end user to set up a view space and add objects to the environment on which several internal & external forces are applied.

This specification documents the non-functional requirements of the Soft-body system like performance, usability, and reliability, to name a few. Also, it defines functional requirements defined in the use cases.

### 9.1.3 Overview

This document is written according the "Dynamic Deformation of Uniform Elastic Two-Layer Objects" thesis whose author is Ms. Miao Song. The rest of the Supplementary Specification will address the functional requirements of the system that address functionality, usability, reliability, performance, and supportability. Then, the document will address design constraints on the system being built. The Online User Documentation and Help System Requirements will be discussed followed by a list of the interfaces that must be supported by the application. Finally, we will address applicable standards applied on the system and the glossary.

## 9.2 Functionality

This section lists Soft-body system functional requirements retrieved from the system's use case model.

### 9.2.1 Save a Simulation

The user chooses to save the simulation by clicking on "start save simulation" button. As a consequence, the system monitors and keeps track of objects attributes: position, applied forces, velocity and mass. When the user clicks on "Stop Save", an XML file format is created and saved in the application default directory.

### 9.2.2 Drag Object

This use case describes end-user interaction with the system using a mouse. The user drags an object and releases it to create a force. As a result, the system shall find the nearest particle to the current position of the mouse and apply this force on rest particles, which is passed through by springs.

## 9.3 Usability

There are a couple of requirements that has a direct impact on the Soft-body system, as listed below:

### 9.3.1Online help

Online help manual and hard copy documentation shall be available for end-users in order to become more productive and use the full set of the features provided by the system

### 9.3.2 IDE Development tools

The Soft-body system shall be developed with a language that can be deployed into different platforms.

### 9.3.3 Graphical Libraries Use

The system uses graphical libraries like OpenGL, GLSL, GLUI, Direct X in order to support the multiple dimensions of the view space: 1D, 2D, and 3D.

## 9.4 Reliability

This section lists the Soft-body system reliability requirements as follow:

### 9.4.1 Availability

The Soft-body system shall be available and operational 24 hours a day during all weekdays. There shall be a minimum down time of 3 % due to system maintenance.

### 9.4.2 Mean Time Between Failures (MTBF)

The Mean Time Between Failures shall exceed 420 hours.

### 9.4.3 Mean Time To Repair

The Mean Time to repair shall not exceed 8 hours.

### 9.4.4 Accuracy

A uniform modeling technique has to be used for surface refinement to generate approximations to curves and surfaces of a sphere.

### 9.4.5 Maximum bugs or defects rate

The maximum defect rate that the application could support is 3% bugs/KLOC (thousands of lines of code).

### 9.4.6 Bugs or defects rate

A minor Soft-body defect shall deviate the behavior of the system from what is expected while a critical defect shall cause the inability to use all the features of the system or leads to a system crash.

## 9.5 Performance

This section outlines the performance characteristics of the Soft-body system:

### 9.5.1 Response Time for a Transaction

The Soft-body system shall have a fast response time allowing 90% of all user transactions to get executed with the interval of 1.2 minutes.

### 9.5.2 Throughput

The Soft-body system shall have the average rate of successful message delivery over the physical link of 50 bits/s

### 9.5.3 Capacity

The Soft-body system shall support only 1 user at a time. In future releases, the system should be distributed over multiple workstations enabling multiple usages.

### 9.5.4 Database Access Response Time

Soft-body system shall interact with a central database system allowing an access with no more than 8 seconds latency.

## 9.6 Supportability

This section indicates the set of requirements that will enhance the supportability or maintainability of the system being built

### 9.6.1 Graphical Libraries

To aid rendering computer graphics to the monitor, graphical libraries like OpenGL, GLSL, GLUI, and Direct X shall be used to handle Soft-body system rendering tasks.

## 9.7 Design Constraints

This section should indicate any design constraints on the system being built

### 9.7.1 Software Languages

The Soft-body system shall support multiple languages: French, English, Spanish, Chinese, and Korean. Therefore, the System shall detect the current workstation culture and load the appropriate language resource library to display the corresponding strings to the user. To achieve this goal, every string should be translated to multiple languages in the language resource file.

### 9.7.2 Software Legacy system

The Soft-body system shall interoperate with existing legacy libraries in order to populate the Objects toolbox with legacy custom objects (particles, springs, faces, and integrators)

### 9.7.3 Platform requirements

The computer deploying the Soft-body system shall have at least 4 GB RAM in order to support multiple graphical libraries used and at least 120GB disk space. The platform shall operate with a SQL server database management system to store, modify, and retrieve data.

## 9.8 Online User Documentation and Help System Requirements

The Soft-body system provides online documentation in order to familiarize end-users with the system and properly use its full features. The Help button shall be located in the menu bar. After the mouse click, a web page navigator should be opened and should load the corresponding help ID from the SQL database. The help menu window shall contain a list of all the main topics in ascending alphabetical order.

## 9.9 Purchased Components

In order to respect the implementation standards like naming and coding conventions and ensure code inspections, refactoring, and fast navigation, an IDE's plug-in should be used: CodeEnhance. This tool requires a license to enable software developers to use the full set of features it provides.

## 9.10 Interfaces

This section defines the interfaces that must be supported by the application.

### 9.10.1 User Interfaces

Soft-body user interfaces shall be interactive and user friendly. The GUI shall adopt eye-friendly colors and shall contain a unified menu bar accessible at the top side of the application. Also, the application shall contain a toolbox with all the objects that the end-user could drag to the view space. There shall be a side menu at the right hand side of the screen with a green Start button to start the simulation.

### 9.10.2 Hardware Interfaces
The Concordia Local Area Network shall be used in order to communicate with the central database server.

### 9.10.3 Software Interfaces
The Soft-body system shall be used with lightweight SQL version Database management System that could be installed on Concordia old workstation that have limited memory storage.

### 9.10.4 Communications Interfaces
The Soft-body system shall communicate with legacy applications through the Local Area Network

## 9.11 Licensing Requirements
At a first step of Soft-body system delivery, the system usage is restricted to Concordia university students to collect feedbacks in order to improve the system before launching it to the market.

## 9.12 Legal, Copyright and Other Notices
The Soft-body system is a trademark of Concordia University and shall not be copied or used without the university permission

## 9.13 Applicable Standards
A couple of criteria shall be applied to the Soft-body system in order to follow the international Organization for Standardization ISO9000 standards. This will address various aspects of quality management to provide guidance so as to ensure that the final product meets the client's requirements with a focus on quality improvement.

# 10. Use Cases

## 10.1 Use Case Briefs

### 10.1.1 Drag Object

This use case describes end-user interaction with the system using a mouse. The user drags an object and releases it to create a force. As a result, the system shall find the nearest particle to the current position of the mouse and apply this force on rest particles, which is passed through by springs.

### 10.1.2 Save simulation

The user chooses to save the simulation by clicking on "start save simulation" button. As a consequence, the system monitors and keeps track of objects' attributes: position, all forces, velocity and mass. When the user clicks on "Stop Save" button, the system pops up a dialog window to the user whether s/he wants to save the simulation. If yes, the system saves the simulation in the application default directory with a default name.

### 10.1.3 Calculate Total Force

The system should calculate the total force applied on the object in order to correctly simulate the collision impact. This force in is fact a sum of different other forces that the object under study is exposed to. These forces could be summarized to: External, gravity, mouse, spring, pressure, and collisions detection forces. The sum of all these forces will provide the total force applied to the object to cause deformation.

### 10.1.4 Link Object

The system shall allow Soft-body objects to be attached to other objects. The user shall set his view space and add objects to it. The objects could be a particle, spring, face, and/or integrator. Then, the application should be able to attach sensors on actors' bodies and link between them to form a user's setting to be used for simulation.

### 10.1.5 Change Object Dimension

The user selects the desired dimension from drop down list. Then, the system shall locate the dimension libraries. The user chooses a dimension that triggers the corresponding library to be loaded providing that it is not corrupted and different than the current dimension. Then, the system unloads the current dimension library and loads the desired one and runs it.

### 10.1.6 Process Idle Object Status

The user did not ask for a new simulation for a moment. Then, the system shall update the object position upon to a random position, and when it reaches to that position, the system shall choose another random position. Also, the system shall consider the forces that have been given by the user for the last simulation. The object shall keep moving though the simulation has not started yet.

## 10.2 Use Case Diagram

### 10.2.1 Use Case Context Diagram

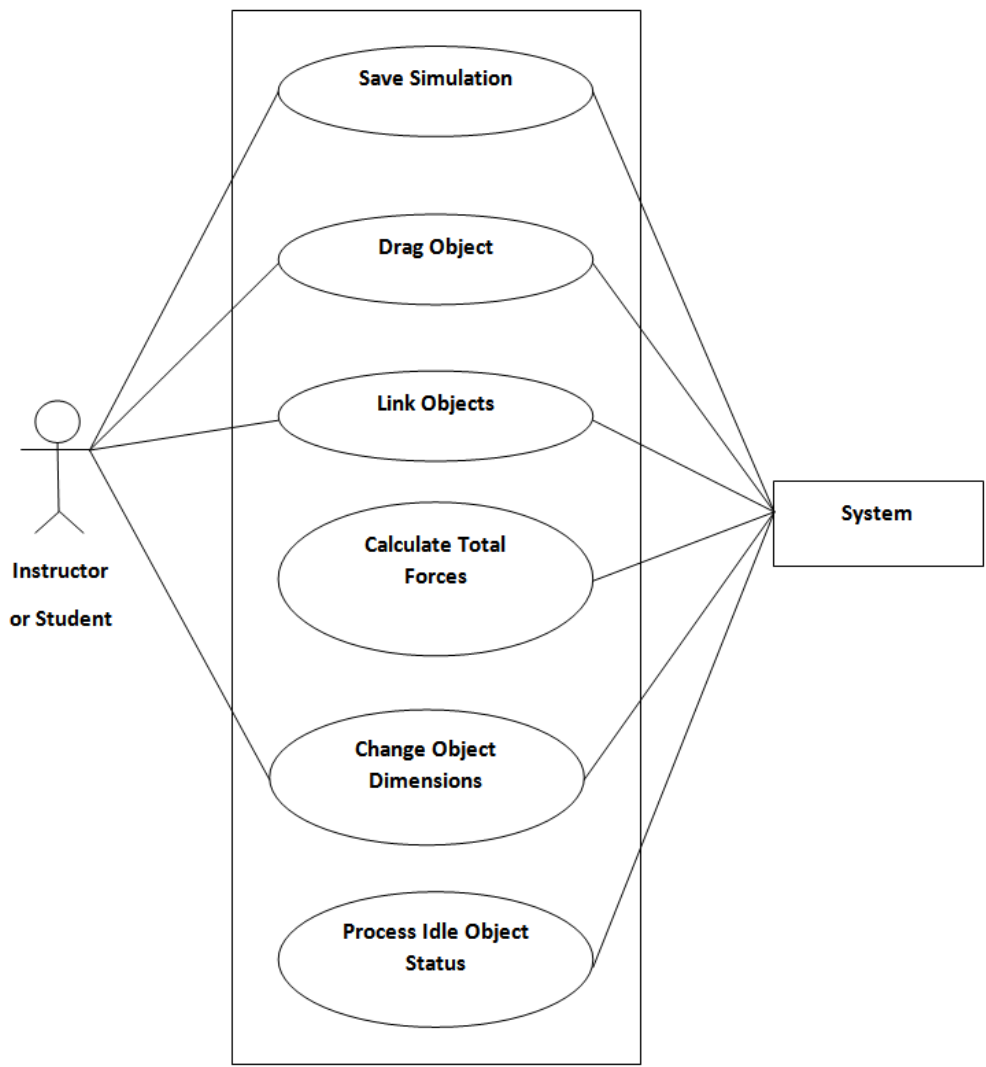

Figure 1 Use Case Context Diagram

### 10.2.2 Drag Object Use Case Diagram

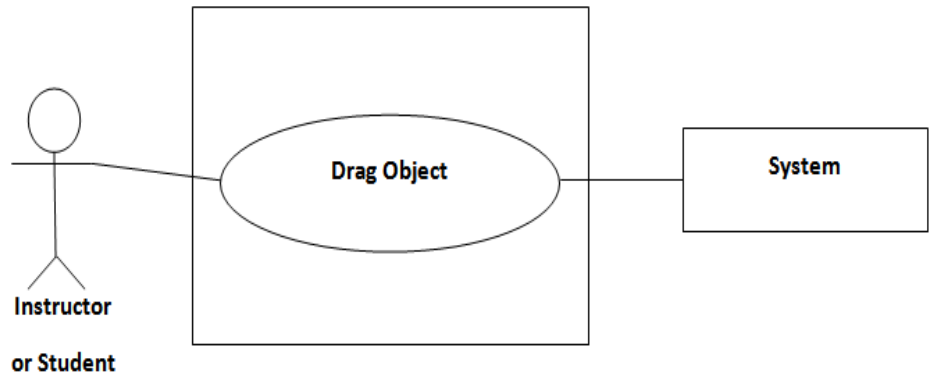

### 10.2.3 Save Simulation Use Case Diagram

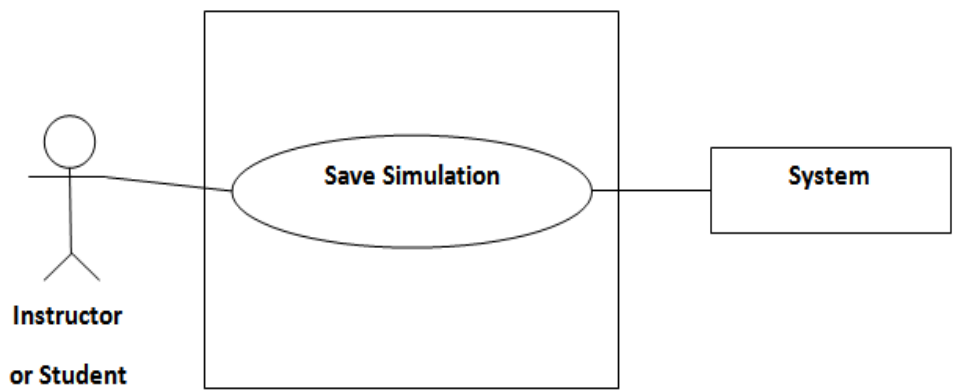

### 10.2.4 Calculate Total Force Use Case Diagram

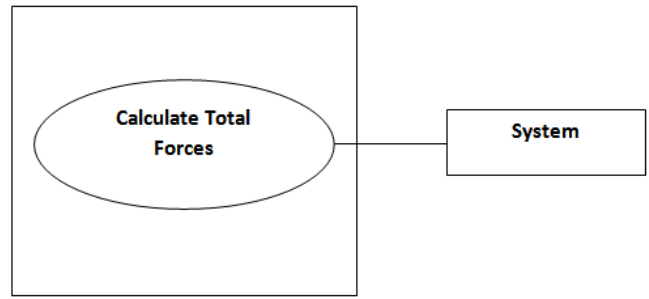

**Figure 4 Calculate Total Force Use Case Diagram**

### 10.2.5 Link Object Use Case Diagram

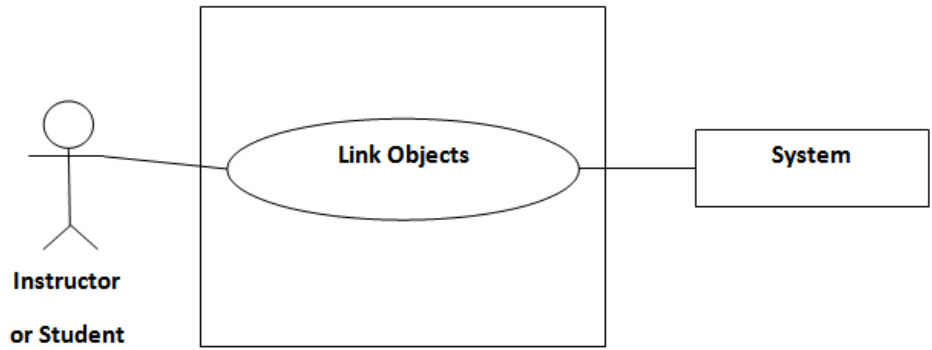

**Figure 5 Link Object Use Case Diagram**

### 10.2.6 Change Object Dimension Use Case Diagram

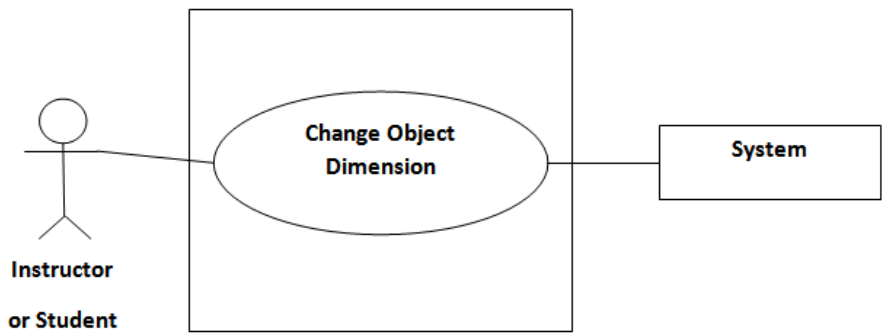

**Figure 6 Change Object Dimension Use Case Diagram**

### 10.2.7 Process Idle Object Status Use Case Diagram

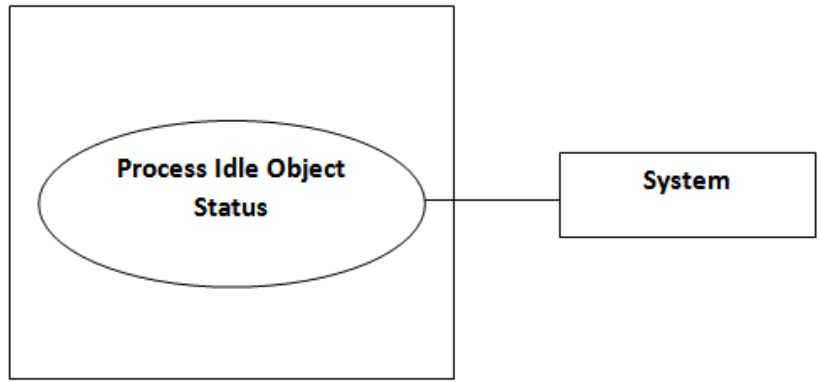

## 10.3 Cost-Value Prioritization for Use Cases

Requirements Prioritization Using Analytic Hierarchy Process (AHP) Method

1 Contributes equally.                    3 Contributes slightly more.

5 Contributes strongly more.              7 Contributes very strongly more.

9 Contributes extremely more.

|  | Drag Object | Save Simulation | Process Idle Object Status | Change Object Dimension | Link Object | Calculate Total Forces |
|---|---|---|---|---|---|---|
| Drag Object | 1 | 5 | 7 | 9 | 7 | 7 |
| Save Simulation | 1 / 5 | 1 | 5 | 9 | 5 | 7 |
| Process Idle Object Status | 1 / 7 | 1 / 5 | 1 | 9 | 5 | 5 |
| Change Object Dimension | 1 / 9 | 1 / 9 | 1 / 9 | 1 | 1 / 9 | 1 / 7 |
| Link Object | 1 / 7 | 1 / 5 | 1 / 5 | 9 | 1 | 5 |
| Calculate Total Forces | 1 / 7 | 1 / 7 | 1 / 5 | 7 | 1 / 5 | 1 |

Table 2 Comparison matrix with relative values of requirements

| | Drag Object | Save Simulation | Process Idle Object Status | Change Object Dimension | Link Object | Calculate Total Forces | Relative Value |
|---|---|---|---|---|---|---|---|
| Drag Object | 0.57 | 0.75 | 0.52 | 0.20 | 0.38 | 0.28 | **0.45** |
| Save Simulation | 0.11 | 0.15 | 0.37 | 0.20 | 0.27 | 0.28 | **0.23** |
| Process Idle Object Status | 0.08 | 0.03 | 0.07 | 0.20 | 0.27 | 0.19 | **0.14** |
| Change Object Dimension | 0.06 | 0.02 | 0.01 | 0.02 | 0.01 | 0.01 | **0.02** |
| Link Object | 0.08 | 0.03 | 0.01 | 0.20 | 0.05 | 0.19 | **0.09** |
| Calculate Total Forces | 0.08 | 0.02 | 0.01 | 0.16 | 0.01 | 0.04 | **0.05** |

Table 3 The normalized matrix and relative contribution of requirements to the project's overall value

| | Drag Object | Save Simulation | Process Idle Object Status | Change Object Dimension | Link Object | Calculate Total Forces |
|---|---|---|---|---|---|---|
| Drag Object | 1 | 5 | 5 | 9 | 7 | 7 |
| Save Simulation | 1 / 5 | 1 | 5 | 9 | 5 | 5 |
| Process Idle Object Status | 1 / 5 | 1 / 5 | 1 | 9 | 5 | 7 |
| Change Object Dimension | 1 / 9 | 1 / 9 | 1 / 9 | 1 | 1 / 9 | 1 / 9 |
| Link Object | 1 / 7 | 1 / 5 | 1 / 5 | 9 | 1 | 7 |
| Calculate Total Forces | 1 / 7 | 1 / 5 | 1 / 7 | 9 | 1 / 7 | 1 |

Table 4 Comparison matrix with relative costs of requirements

| | Drag Object | Save Simulation | Process Idle Object Status | Change Object Dimension | Link Object | Calculate Total Forces | **Relative Cost** |
|---|---|---|---|---|---|---|---|
| Drag Object | 0.56 | 0.57 | 0.44 | 0.19 | 0.38 | 0.26 | **0.40** |
| Save Simulation | 0.11 | 0.15 | 0.44 | 0.19 | 0.27 | 0.18 | **0.22** |
| Process Idle Object Status | 0.11 | 0.03 | 0.09 | 0.19 | 0.27 | 0.26 | **0.16** |
| Change Object Dimension | 0.06 | 0.02 | 0.01 | 0.02 | 0.01 | 0.01 | **0.02** |
| Link Object | 0.08 | 0.03 | 0.02 | 0.19 | 0.05 | 0.26 | **0.11** |
| Calculate Total Forces | 0.08 | 0.03 | 0.01 | 0.19 | 0.01 | 0.04 | **0.06** |

Table 5 The normalized matrix and relative contribution of requirements to the project's overall cost

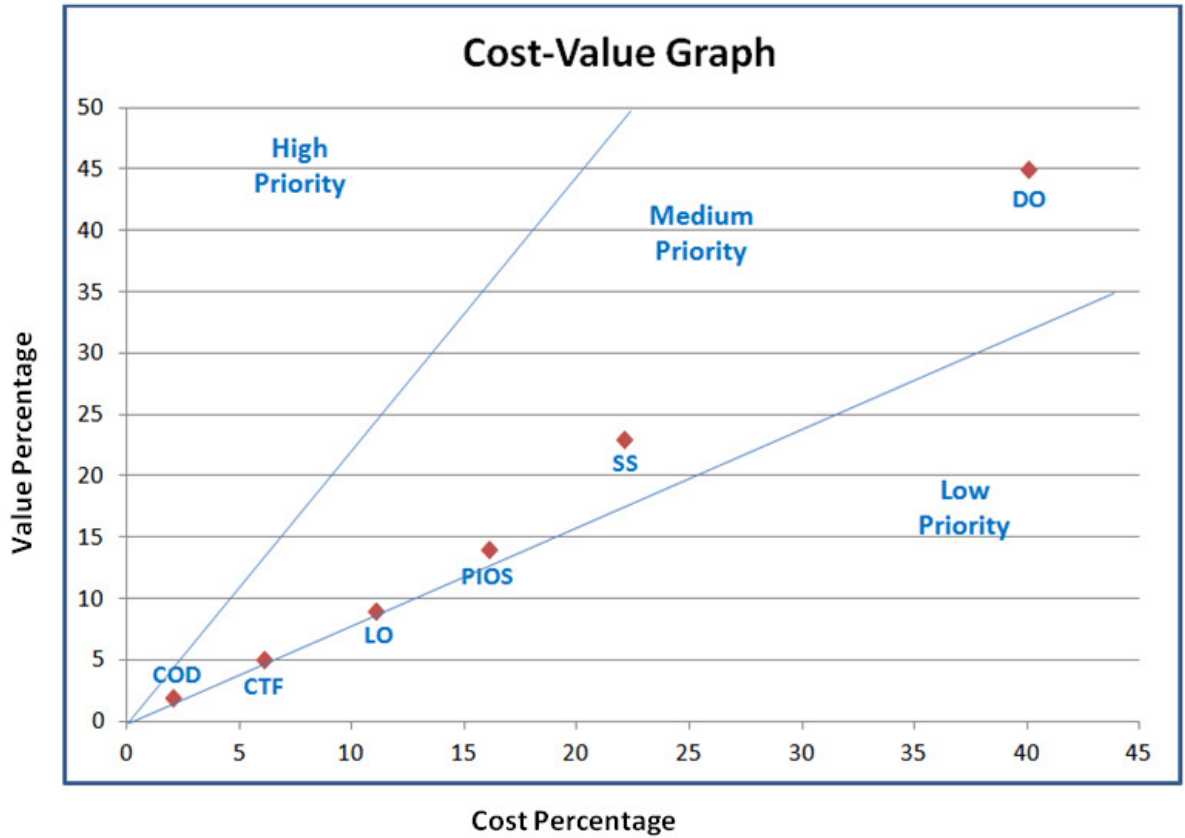

Figure 8 Cost-Value requirements prioritization obtained by AHP method

DO: Drag Object                         SS: Save Simulation

PIOS: Process Idle Object Status        COD: Change Object Dimension

LO: Link Object                         CTF: Calculate Total Forces

## 10.4 Fully Dressed Use Cases

### 10.4.1 Drag Object

| Use Case ID | UC_DragObject |
|---|---|
| Use case name | Drag Object |
| Scope | System under investigation |
| Level | User level |
| Primary Actor | Instructors / Students |
| Stakeholders and Interests | Student/Instructor: Want a quick, accurate and real time effect that reflects normal laws of physics. |
| Preconditions | • There is a mouse connected to the computer.<br><br>• The user has set the Soft body object properties.<br><br>• There must be at least one external force applied on the object: (gravity, mouse, spring, pressure, or collisions detection force). |
| Post conditions | The Soft body object is dragged accordingly. |
| Main success scenario | 1. User indicates to the system that s/he tends to start a simulation.<br><br>2. User clicks on the soft body object and holds it with the mouse click.<br><br>3. While holding the click, user moves the mouse and sees its cursor and perceives the object movements.<br><br>4. The system shall find the nearest particle to the current position of the mouse.<br><br>5. The new user applied force shall affect the forces applied on rest particles.<br><br>6. User releases the mouse click.<br><br>7. The system keeps updating the soft body object's position according to the applied forces. |
| Extensions | Shall the system crash at any time, the user needs to close and reopen it. |
| | 3a: The user drags the soft body object using keyboard's arrows. |
| |     1- The user uses the arrows to drag the soft body object to the direction s/he wants. |
| |     2- Flow goes to 4 in main scenario. |
| | 6a: The user drags the object until it reaches the view space border. |
| |     1- The system stops updating the object's position. |
| |     2- The user releases the mouse. |
| |     3- Flow goes to 7 in main scenario. |
| |         6a: 2a: The user drags the object back to the view space. |
| |     1- Flow goes to 3 in main scenario. |
| Special requirements | The view space should be interactive and the graphical user interface should support 3D view. |
| Technology and data variations list | The drag force applied on the object creates a force whose magnitude is displayed to the user in the following format: X.XXXX (Four digits after the dot) |
| Frequency of occurrence | The dragging process should take place anytime the user drags a given object |
| Miscellaneous | What happens when a crash occurs due to the dragging process?<br><br>A log file will be kept containing a call stack with the current exception to help developers to fix the defect. |

### 10.4.2 Save Simulation

| Use Case ID | UC_SaveSimulation |
|---|---|
| Use case name | Save a simulation |
| Scope | System under investigation |
| Level | User level |
| Primary Actor | Instructors / Students |
| Stakeholders and Interests | Instructors / Students: Want correct and fast process to save a simulation. |
| Preconditions | The simulation is running. |
| Post conditions | The simulation is correctly saved. |
| Main success scenario | 1. The user clicks on the "start save simulation" button.<br>2. The system monitors and keeps track of objects' attributes: position, all forces, velocity and mass.<br>3. The user clicks on "stop saving" button to stop saving.<br>4. The system pops up a dialog window that has a file default name for the simulation and the application default directory where to save it for the user to confirm.<br>5. The user clicks on save.<br>6. The system saves the simulation. |
| Extensions | Shall the system crash at any time, the user needs to close and reopen it.<br><br>2a: The memory is full.<br><br>    1- The system stops the saving process.<br><br>    2- The system informs the user of the error. |

| | 3- Flow goes to 4 in main scenario. |
|---|---|
| | 4a: The user wants to save the file in another file directory. |
| |     1- The user clicks on browse button from the pop up window. |
| |     2- The user chooses a folder and clicks on Save. |
| |     3- Flow goes to 6 in the main scenario. |
| | 4b: The user gives the simulation file a name different than the default one. |
| |     1- The user enters the file's name in the file name field in the pop up window. |
| |     2- Flow goes to 5 in the main scenario. |
| Special requirements | The system monitors and tracks objects' attributes in real time. |
| Technology and data variations list | The file format of the saved file could be an XML or CSV file. |
| Frequency of occurrence | Could be nearly continuous. |
| Miscellaneous | It needs a good recovery support, what happens if the system crashes while the simulation is being saved? |

## 11. Domain Model

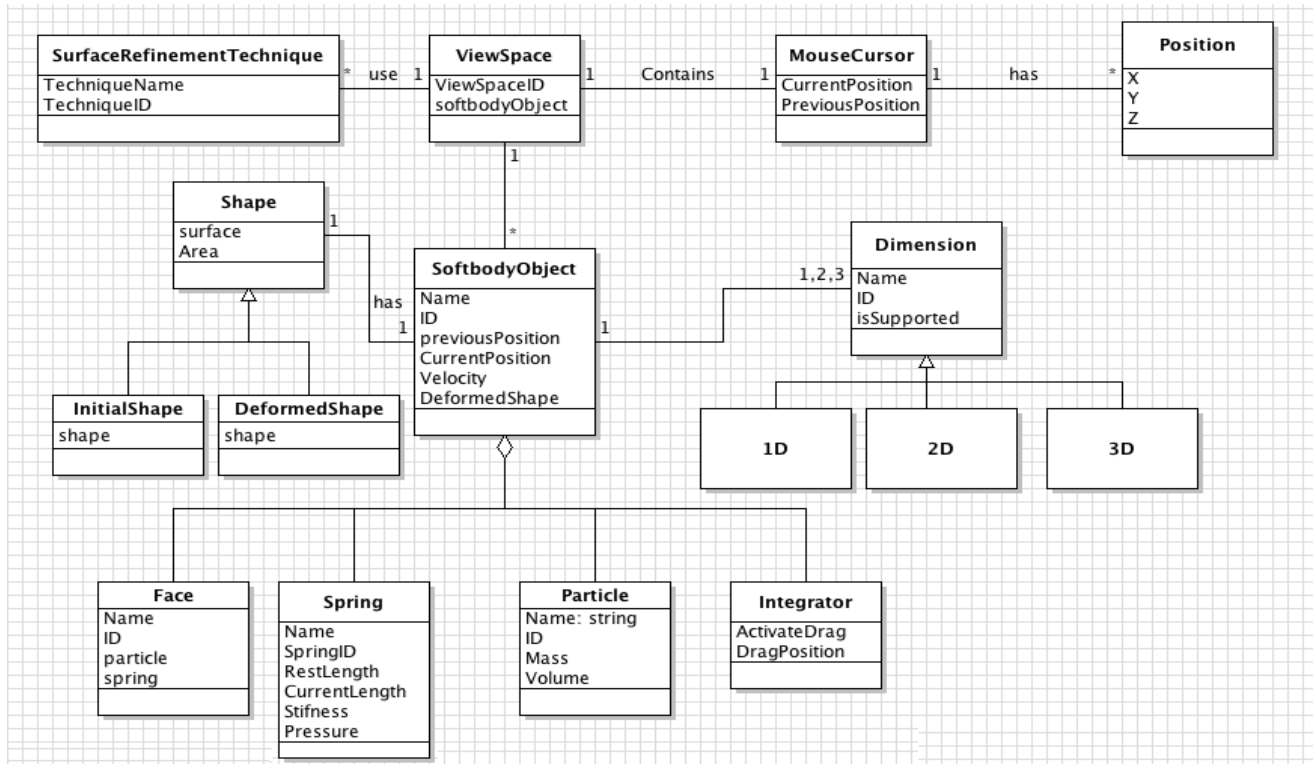

Figure 9 Domain Model Diagram

# 12. System Sequence Diagram

## 12.1 Drag Object

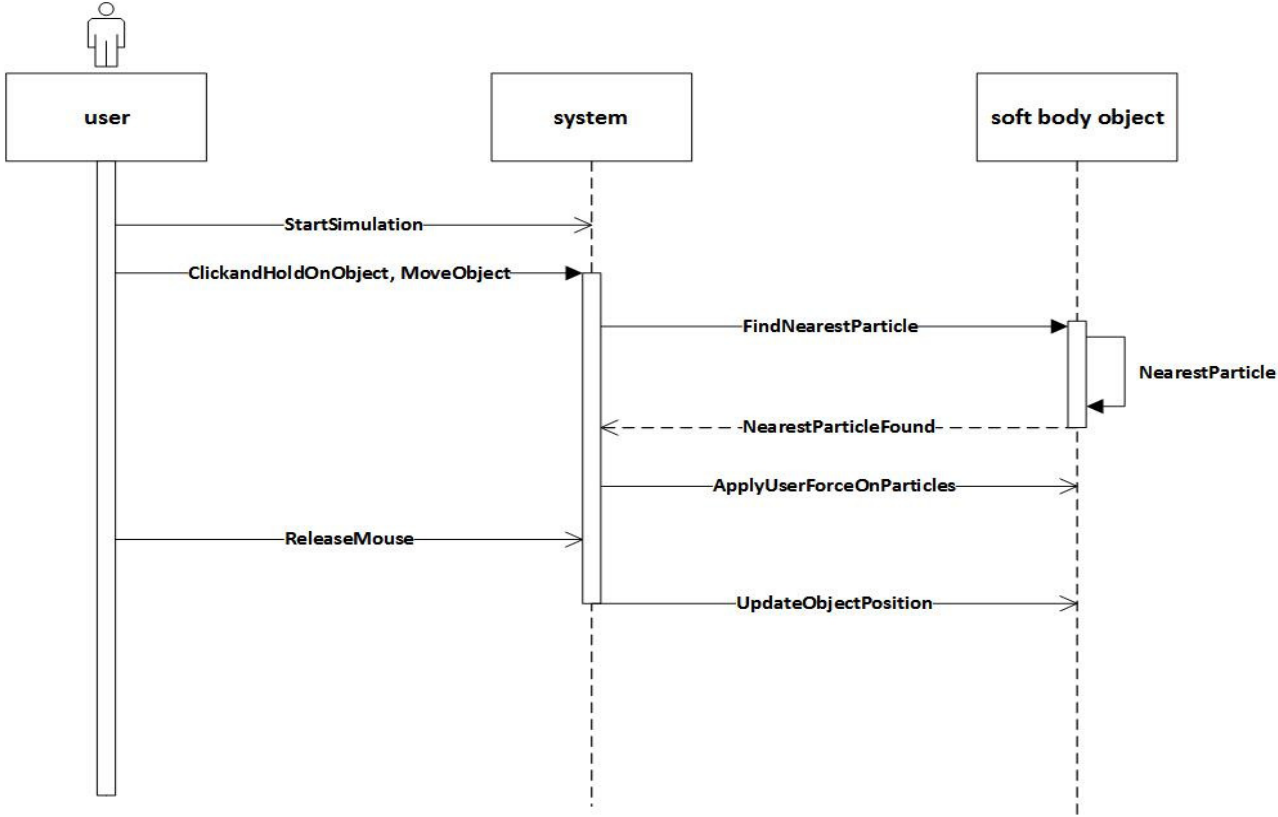

**Figure 10 SSD for Drag Object Use Case**

## 12.2 Save Simulation

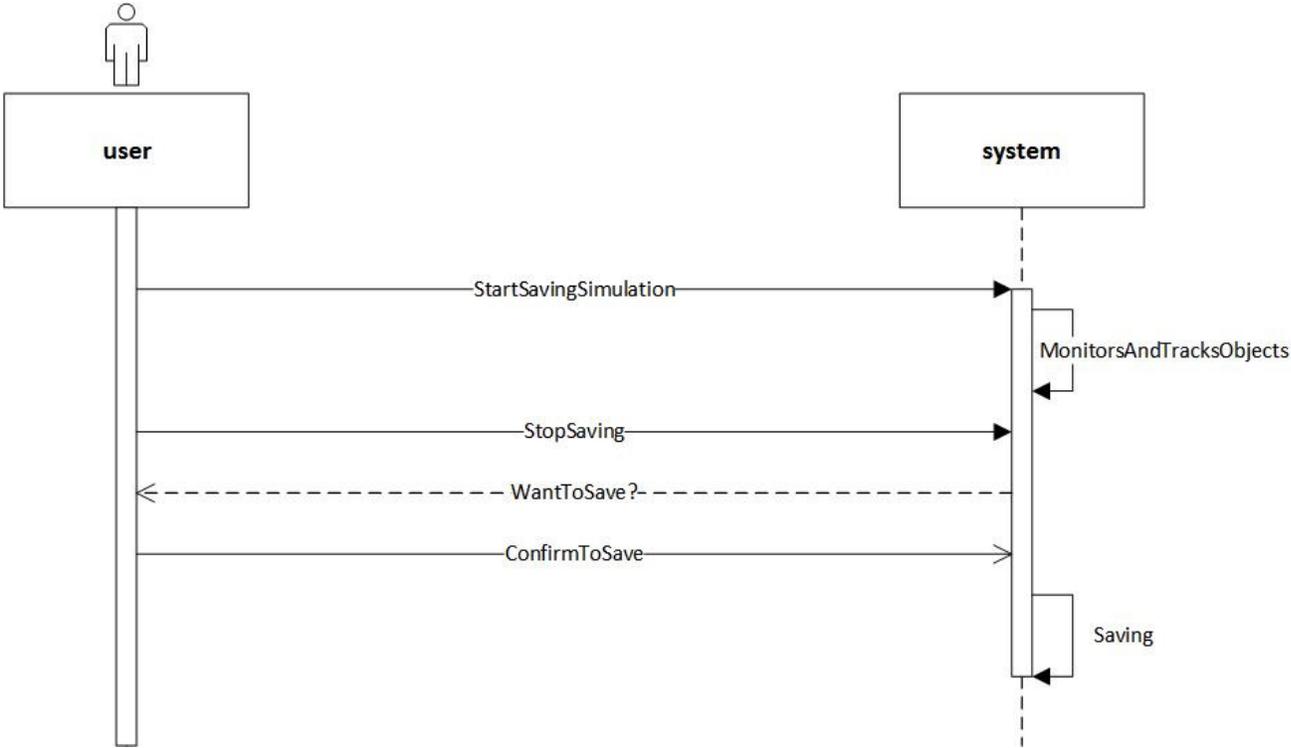

**Figure 11 SSD for Save Simulation Use Case**

# 13. Activity Diagram

## 13.1 Drag Object

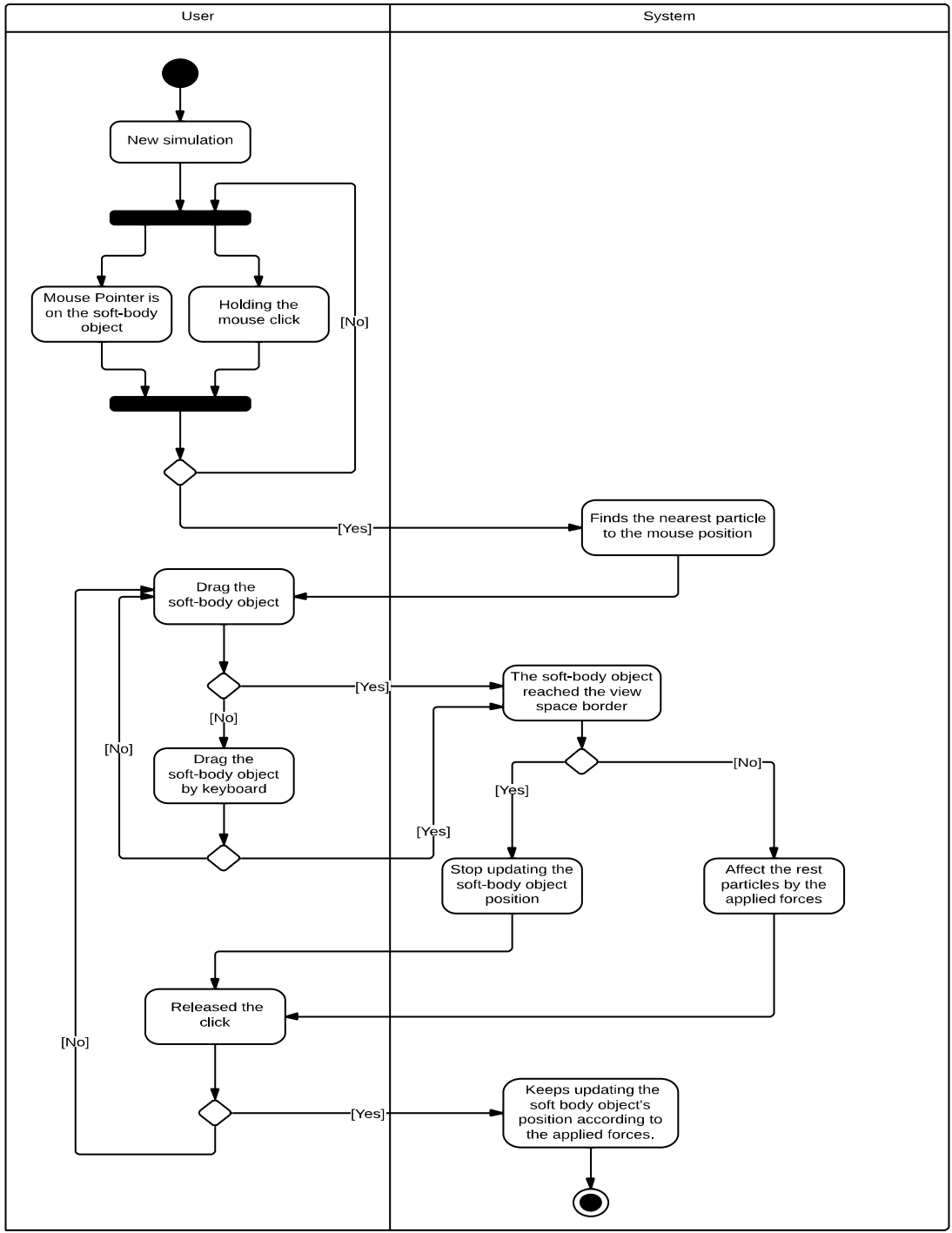

Figure 12 Activity Diagram for Drag Object Use Case

## 13.2 Save Simulation

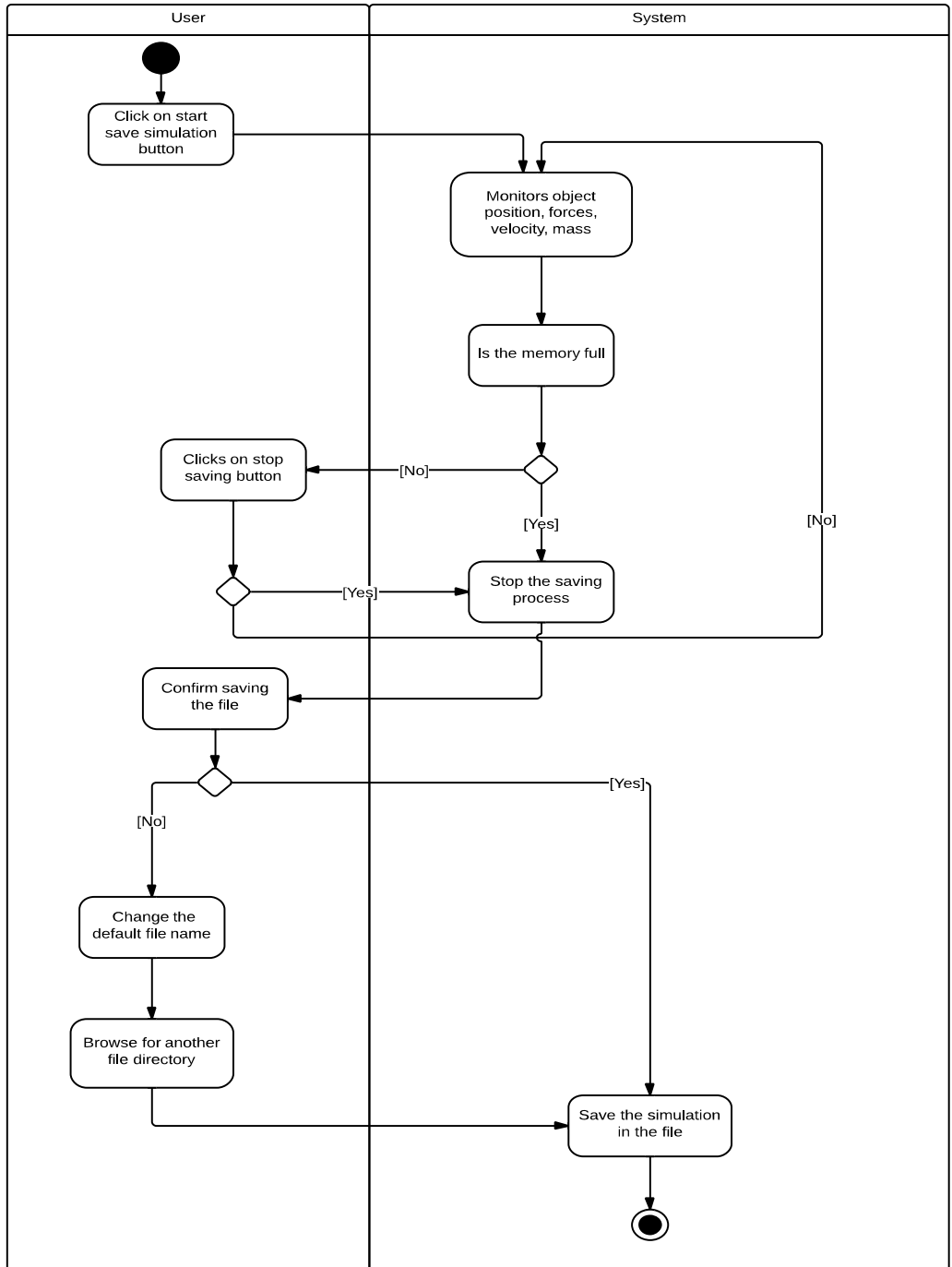

**Figure 13 Activity Diagram for Save Simulation Use Case**

# 14. Test Cases

| Test Case ID | TC_DragObject_1 |
|---|---|
| Title | Test Case for UC_DragObject, Main Scenario |
| Requirement | Main scenario of UC_DragObject |
| Type | Acceptance |
| Settings | • Graphics Card that Supports 3D.<br>• Screen that supports 3D.<br>• There is a mouse connected to the computer.<br>• There is a keyboard connected to the computer. |
| Preconditions | • Objects' attributes are set.<br>• User started the simulation. |
| Description | The user drags the soft body object as described in the use case's main scenario. |
| Expected Results | The soft body object moves according to the dragging direction and strength. |

<p align="center"><strong>Table 6 Test Case TC_DragObject_1</strong></p>

| Test Case ID | TC_DragObject_2 |
|---|---|
| Title | Test Case for UC_DragObject, user drags the soft body object using keyboard's arrows. |
| Requirement | Scenario 3a in UC_DragObject |
| Type | Acceptance |
| Settings | • Graphics Card that Supports 3D. |

|  | • Screen that supports 3D. |
|  | • There is a mouse connected to the computer. |
|  | • There is a keyboard connected to the computer. |
| **Preconditions** | • Objects' attributes are set. |
|  | • User started the simulation. |
| **Description** | The user clicks and hold the soft body object using a mouse, then drags it using the keyboard's arrows. |
| **Expected Results** | The soft body object moves according to the keyboard's direction arrows. |

**Table 7 Test Case TC_DragObject_2**

| **Test Case ID** | TC_DragObject_3 |
| **Title** | Test Case for UC_DragObject, user drags the soft body object to the view border. |
| **Requirement** | Scenario 6a in UC_DragObject |
| **Type** | Acceptance |
| **Settings** | • Graphics Card that Supports 3D. |
|  | • Screen that supports 3D. |
|  | • There is a mouse connected to the computer. |
|  | • There is a keyboard connected to the computer. |
| **Preconditions** | • Objects' attributes are set. |
|  | • User started the simulation. |
| **Description** | The user drags the soft body object to the view border and releases the mouse click. |
| **Expected Results** | The system updates the soft body object position. |
| **Test Case ID** | TC_SaveSimulation_1 |
| **Title** | Test Case for UC_SaveSimulation, main scenario |
| **Requirement** | Main scenario of UC_SaveSimulation |
| **Type** | Acceptance |
| **Settings** | • Graphics Card that Supports 3D.<br>• Screen that supports 3D. |
| **Preconditions** | User started the simulation. |
| **Description** | The user starts the simulation then clicks save button, the system monitors objects' attributes. When the user clicks stop button, the system saves the simulation. |
| **Expected Results** | • The system saves the simulation file in the application default directory.<br>• The simulation file's name is the application default name. |
| **Test Case ID** | TC_SaveSimulation_2 |
| **Title** | Test Case for UC_SaveSimulation, the system's memory is full. |
| **Requirement** | Scenario 2a of UC_SaveSimulation |
| **Type** | Acceptance |
| **Settings** | • Graphics Card that Supports 3D.<br>• Screen that supports 3D. |

| Preconditions | User started the simulation. |
|---|---|
| Description | The user starts the simulation then clicks save button, the system monitors objects' attributes, and the system's memory becomes full. Then the system stops the simulation saving process. |
| Expected Results | The system stops the simulation saving process, notifies the user of the error and pops up the saving window. |

<div align="center">Table 10 Test Case TC_SaveSimulation_2</div>

| Test Case ID | TC_SaveSimulation_3 |
|---|---|
| Title | Test Case for UC_SaveSimulation, user chooses a file directory where to save the simulation file. |
| Requirement | Scenario 4a in UC_SaveSimulation |
| Type | Acceptance |
| Settings | • Graphics Card that Supports 3D.<br>• Screen that supports 3D. |
| Preconditions | User started the simulation. |
| Description | User starts the simulation then clicks save button, system monitors objects' attributes. The user clicks stop button and chooses a file directory to save the simulation file in it. |
| Expected Results | • The system saves the simulation file in the user chosen file directory.<br>• The simulation file's name is the application default name. |

<div align="center">Table 11 Test Case TC_SaveSimulation_3</div>

| Test Case ID | TC_SaveSimulation_4 |
|---|---|
| Title | Test Case for UC_SaveSimulation, user gives the simulation file a name. |
| Requirement | Scenario 4b in UC_SaveSimulation |
| Type | Acceptance |
| Settings | • Graphics Card that Supports 3D.<br>• Screen that supports 3D. |
| Preconditions | User started the simulation. |
| Description | User starts the simulation then clicks save button, system monitors objects' attributes. The user clicks stop button and gives a name to the simulation file. |
| Expected Results | • The system saves the simulation file in the application default directory.<br>• The simulation file's name is that the user entered. |

<div align="center">Table 12 Test Case TC_SaveSimulation_4</div>

# 15. Traceability Matrices

## 15.1 User Needs versus Features

|  | F1 | F2 | F3 | F4 | F5 | F6 | F7 | F8 | F9 | F10 |
|---|---|---|---|---|---|---|---|---|---|---|
| N1 | X |  |  |  |  | X |  |  |  | X |
| N2 | X |  |  | X |  | X |  |  |  | X |
| N3 |  |  |  |  | X |  |  |  |  |  |
| N4 |  |  |  | X |  | X |  |  |  | X |
| N5 |  | X |  |  |  |  | X | X |  |  |

| | F1 | F2 | F3 | F4 | F5 | F6 | F7 | F8 | F9 | F10 |
|----|----|----|----|----|----|----|----|----|----|----|
| N6 | | | X | | | | | | | |
| N7 | X | | | | | | | | | X |
| N8 | | | | | | | | | X | |

<div align="center">Table 13 User Needs versus Features</div>

**User Needs:**

N1:  Interact with the object at real time.

N2:  Squash and stretch the object according to external and internal forces

N3:  The object shall be simulated in different views (1,2 or 3D)

N4:  The object shall not stop and keeps moving as if it is a live Jellyfish

N5:  Save the simulation at any time

N6:  Link objects to other objects

N7:  Drag the object by mouse, Keyboard and/or optic device.

N8:  Initiate the object with desired properties

**Features:**

F1:   Interact with the object by mouse

F2:   Save the simulation into a file

F3:   Link objects to each other

F4:   Support idle state for the object

F5:   Select object dimension.

F6:   Apply internal / external forces on the object

F7:   Save the file in a default application directory or a different one

F8:   Stop saving process

F9:   Set object properties

F10:  Update object position

## 15.2 Features versus Use Cases

| | UC_DragObject | UC_SaveSimulation | UC_LinkObject | UC_ProcessIdleObjectState | UC_CalculateTotalForces | UC_ChangeObjectDimension |
|---|---|---|---|---|---|---|
| F1 | X | X | X | | | |
| F2 | | X | | | | |
| F3 | | | X | | | |
| F4 | | | | X | | |
| F5 | | | | | | X |
| F6 | | | | | X | |
| F7 | | X | | | | |
| F8 | | X | | | | |
| F9 | | | | | | X |
| F10 | X | | | X | | |

**Table 14 Features versus Use Cases**

UC_DragObject: Drag Object

UC_ SaveSimulation: Save Simulation

UC_LinkObject: Link object

UC_ ProcessIdleObjectState: Process Idle Object State

UC_ CalculateTotalForces: Calculate Total Forces

UC_ ChangeObjectDimension: Change Object Dimension

## 15.3 Features versus Supplementary Requirements

| | Functionality | Usability | Reliability | Performance | Supportability | Design Constraints | Interfaces |
|---|---|---|---|---|---|---|---|
| Drag Object | X | | | | | | |
| Save Simulation | X | | | | | | |
| Online Help | | X | | | | | |
| IDE Development tools | | X | | | | | |
| Graphical Libraries Use | | X | | | | | |
| Availability | | | X | | | | |
| Mean Time Between Failures (MTBF) | | | X | | | | |
| Mean Time To Repair | | | X | | | | |
| Accuracy | | | X | | | | |
| Maximum bugs or defects rate | | | X | | | | |
| Bugs or defects rate | | | X | | | | |
| Response Time for a Transaction | | | | X | | | |
| Throughput | | | | X | | | |
| Capacity | | | | X | | | |
| Database Access Response Time | | | | X | | | |
| Graphical Libraries | | | | | X | | |
| Software Languages | | | | | | X | |
| Software Legacy System | | | | | | X | |
| Platform requirements | | | | | | X | |
| User Interfaces | | | | | | | X |
| Hardware Interfaces | | | | | | | X |
| Software Interfaces | | | | | | | X |
| Communication Interfaces | | | | | | | X |

<div align="center"><span style="color:#4472C4">**Table 15 Features versus Supplementary Requirements**</span></div>

## 15.4 Traceability Matrix for Use Cases to Test Cases

| Use Case ID | Scenario Number | Test Case ID |
|---|---|---|
| UC_DragObject | 1 | TC_DragObject_1 |
| UC_DragObject | 3a | TC_DragObject_2 |
| UC_DragObject | 6a | TC_DragObject_3 |
| UC_SaveSimulation | 1 | TC_SaveSimulation_1 |
| UC_SaveSimulation | 2a | TC_SaveSimulation_2 |
| UC_SaveSimulation | 4a | TC_SaveSimulation_3 |
| UC_SaveSimulation | 4b | TC_SaveSimulation_4 |

<div align="center"><span style="color:#4472C4">**Table 16 Traceability Matrix for Use Cases to Test Cases**</span></div>

# 16. Appendix

## 16.1 Interview

**Part I: Establishing the Customer of User Profile**

Name: Miao Song

Company: Department of Computer Science & Engineering, Concordia University

Industry: Research & Development

Job Title: Research Assistant

**What are your key Responsibilities?**

Key responsibilities is to gather system requirements, analyze, elicit, and build a complete software requirement specification for a virtual jellyfish product.

**What outputs do you produce?**

The output is a functional, reliable, software system that could simulate a real life behavior of jellyfish.

**For Whom?**

- Supervisor: Dr. Peter Grogono.

 - Chair of Department or Graduate Program Director: Dr. Nematollaah Shiri.

- School of Graduate Studies, Concordia University.

**How is success measured?**

Success is measured by the ability to build a software system that could simulate deformation based on the realistic physical consistency of tissues and the laws of the established physics. Success also is measured by ending up with a project that is approved by stakeholders (Supervisors, Chair, and examiner community).

**What are the problems that could interfere with Success?**

- Elastic object simulation is not accurate.
- Software objects are not squashed and stretched according to external & internal forces applied to them.
- System fails to update the object position upon the mouse position.
- System fails to attach Soft-body objects to other objects.
- Graphical user interface does not provide a user interactive environment.
- The collision detection & response is not handled properly.
- The system graphical user interface does not allow the user to drag objects.

**What makes building the system easier and/or difficult?**

The client wants a system that simulates the real-life object deformation according to the established physical laws where user could interact with a system in real-time with an interactive graphical user interface. Also, the client requires that the program should be easy in implementation, and give best elastic body behavior at the minimum cost.

As a matter of fact, a system that simulates an elastic body and real life deformation is an expensive system that requires the integration of latest visualization technologies and precise real-time algorithms that could simulate the forces applied on a given object and simulate the object's reaction to these forces.

**Part II: Assessing the problem**

**For which problems do you lack good solutions?**

A uniform modeling technique has to be adopted for surface refinement to generate approximations to curves and surfaces of a sphere. This technique can be used to describe the complex behavior of the Soft-body system combined with physical laws; however, it can result with irregular surface stiffness and might cause non-spherical shapes.

**How do you solve it now?**

Our simulation system ignores the described drawback resulting from the uniform sphere modeling method.

**Anything else?**

One of the problems that our system encounters is to simulate the forces for every particle connected to four or six springs.

**How do you solve it now?**

The difference of the forces for every particle either connected to four springs or six springs is not addressed in this work.

**Anything else?**

Users cannot animate elastic objects in real-time because the reality and accuracy of the software always require high-end knowledge of physics, mathematics and heavy computations.

**How do you solve the problem?**

In this work, users can only animate elastic object with kinematic modeling method by setting values through the software interface rather than interact with the object in real time.

**Part III: Understanding the User Environment**

**Who are the users?**

The users are:

- Supervisor and research students' colleagues

- Concordia Physics Department.

- Research students in different departments could base their research topics on the system and improve it.

**What is their educational background?**

The educational background of the system users could differ from:

- Simple users who are using the system just to explore the software ability to simulate elastic objects accurately.
- Educational and research backgrounds to enhance research and simulate objects in real time for innovation purposes.

**What is their computer background?**

Users should have good computer background because they need to have knowledge about how to use this complex software in order to use its advanced functionalities.

**Are users experienced with this type of application?**

Users are not experienced with this type of application. They should refer to the user manual or get a training to learn how to use it efficiently.

**What are your expectations for usability of the product?**

I expect that the system will be used in different contexts due to the fact that the existing systems in the market do not handle some of the physical laws issues that the current system does.

**Which platforms are in use?**

As a first project phase, the system is developed using a Microsoft IDE (Visual Studio Environment). However, in future releases the system will be adapted to run under OSX and Linux platforms.

**What are your plans for future platforms?**

We could run the product on Linux in future as well as on a mobile platform (IPhone & Android).

**What kinds of user help does the system provide?**

The system is user friendly and provides an easy-to-use interactive graphical user interface. There is a user manual available that explains in details how to use the software to exploit its maximum capacity. The code is also well documented to ease code understanding as well as system reuse, extension and maintenance.

**What are your expectations for training time?**

The system-to-be does not require enhanced computer skilled users. A few minutes training would be enough for the end user to start using the product.

**Part IV: Recap for understanding**

You have told me:

- Software objects shall be squashed and stretched according to external & internal forces applied on them.
- The two-layer elastic computer generated object provides more accurate modeling based on the main feature of human tissue.
- Deformation is based on the realistic physical consistency of tissues and the laws of the established physics.
- Program shall be easy in implementation, convenient to re-use, and give best elastic body behavior at the minimum cost.
- Users shall be able to interact with the software body in real-time.
- The collision detection & response must be handled correctly.
- The elastic object simulation has to be accurate.
- The three-dimensional elastic object consists of the same elements of 2D, such as particles, springs, and faces, but extended to z-axis.

- A uniform modeling technique has to be used for surface refinement to generate approximations to curves and surfaces of a sphere.
- The system shall be efficient, accurate, and stable.
- An elastic object consists of a particle, spring, face, and integrator.
- The system's graphical user interface provides the user an interactive environment.
- Users can animate elastic object with Kinematic modeling method by setting values through the software interface.
- Users can drag the mass with the mouse to change object position and direction
- The system's graphical user interface allows the user to drag the object.
- The user shall set his view space and choose the object type:
  - One-dimensional.
  - Two-dimensional.
  - Three-dimensional
- The end user shall choose the integrator type:
  - Eurler
  - Midpoint
  - Runge Kutta
- The end user shall set up the springs:
  - Stiffness variables
  - Damping variables
  - Pressure
- The system shall create an elastic object and add particles, springs and faces to the layout.
- The system shall display the object with:
  - The new position
  - Velocity
  - Deformed shape
- The system integrator shall be able to update the particles velocity and new position based on:
  - Accumulated forces

- External forces
- Gravity forces
- Mouse forces
- Spring forces
- Pressure forces
- Collisions detection

- The system updates the object position upon the mouse position.

- The system shall allow Soft-body objects to be attached to other objects.

- The system shall promote interactivity through haptic devices with the Soft-body feedbacks.

- The user shall save a state dump at any time in a CVS or XML file.

- The user can reload functionality from a saved XML file for the following reasons:
  - Simulation
  - Replay
  - Debugging purposes.

- The XML file could be imported into:
  - Relational database.
  - Excel spreadsheet

- The system allows for multiple rendering back-ends.

- The system allows stereoscopic effects.

- User can apply a drag force to the elastic object through the mouse.

- Users can interact with the modeled objects, deform them, and observe the response to their action in real-time

**Part V: The Analyst's input on the Customer's Problem**

**Which, if any, problems are associated with: (list any needs or additional problems you think should concern the customer or user.)**

- The system uses graphical libraries to simulate the object under multiple dimensions. This would have a serious impact on the system performance and reliability that should be addressed

- **Is this a real problem?**

    Yes, it may affect the performance and reliability of the system-to-be.

- **What are the reasons for this problem?**

    Dealing with graphics would need high performance from the computer that is running it.

- **How do you currently solve the problem?**

    By using a computer with high speed CPU and enough RAM.

- **How would you like to solve the problem?**

    By buying a new computer.

- **How would you rank solving these problems in comparison to other you have mentioned?**

    I would say it is a good solution.

**Part VI: Assessing your solution**

What if you could?

- N/A

How would you rank the importance of these?

- N/A

**Part VII: Assessing the Opportunity**

**Who in your organization needs this application?**

Scientific organizations could use this application in order to simulate elastic objects

**How many of these types of users would use the application?**

Different types of users could use the application. They could be Physics students, aircraft science or any field related to objects collision detection and response.

**How would you value a successful solution?**

A successful solution could bring benefits to the users since it reflects the real life collision process and simulates the object behavior due to the forces exposed on it. It could help science to progress and bring computer-simulated objects into life.

**Part VIII: Assessing the reliability, performance, and support needs**

**What are your expectations for reliability?**

The system is reliable because it provides a correct output based on the object's induced forces and displays the best possible approximation to elastic objects based on the results of physical and mathematical equations.

**What are your expectations for performance?**

There is a tradeoff between quality and performance. In order to achieve quality and simulate objects in 3D, for instance, we need to use graphical libraries like OpenGL, GLSL, GLUI, and Direct X that consume memory and could exhaust the microprocessor, which could lead to some performance issues and some flickering.

**What are your expectations for usability of the product?**

The product can be used all over the world. Based on the platform culture that is being used, the application detects the user's language and uses the appropriate language resource to convert the strings to the equivalent language if supported. Hence, the software could be used in different countries and could support different languages.

**Will you support the product or the others will support it?**

I will support the product myself while I am still in the research lab. However, my lab research candidates will support the product in the future and improve it.

**What are the security requirements?**

Security requirements could be performed at the implementation level. We need to adopt a way to secure our code so that competitors could not decompile our libraries.

**Part IX: Other Requirements**

**Are there any legal, regulatory, or environmental requirements or other standers that must be supported?**

- No, there is nothing else should be supported.

**Can you think of any other requirements we should know about?**

- The system updates the object position upon the mouse position.
- The system shall allow Soft-body objects to be attached to other objects.
- The system shall promote interactivity through haptic devices with the Soft-body feedbacks.
- The system uses graphical libraries like OpenGL, GLSL, GLUI, Direct X
- The system shall create an elastic object and add particles, springs and faces to the layout.
- The system's graphical user interface provides the user an interactive environment.
- The program shall be easy in implementation, convenient to re-use, and give best elastic body behavior at the minimum cost.

**Part X: Wrap-up**

**Are there any other questions I should be asking you?**

 - No, there is not.

**If I need to ask follow-up questions, may I give you a call? Would you be willing to participate in a requirements review?**

- You either send me an email or simply stop by my office on Fridays between 8:30-10:00 pm.

**Part XI: the analyst's summary**

After the interview, and while the data is still fresh in your mind, summarize the three highest-priority needs or problems identified be this user/ customer.

1. Implementing a simulation system that is accurate and represents deformation as a realistic physical consistency of tissues reflecting the laws of the established physics.

2. Software objects shall be squashed and stretched according to external & internal forces applied on them.

3. Program shall be easy in implementation, convenient to re-use, and give best elastic body behavior at the minimum cost.